\documentclass[final,3p,times]{elsarticle}
\usepackage{graphics}
\usepackage{graphicx}
\usepackage{epsfig}
\usepackage{amssymb}

\begin{document}
\begin{frontmatter}


\title{Thermal entanglement properties of $N$-qubit quantum Heisenberg chain in a two-component magnetic field }
\author[]{\"{U}mit Ak{\i}nc{\i}}
\author[]{Erol Vatansever\corref{cor1}$^\ast$}
\cortext[]{Corresponding author. Tel.: +90 3019547; fax: +90 2324534188.} \ead{erol.vatansever@deu.edu.tr}
\author[]{Yusuf Y\"{u}ksel}
\address{Department of Physics, Dokuz Eyl\"{u}l University,
Tr-35160 \.{I}zmir, Turkey}


\begin{abstract}
We elucidate the finite temperature entanglement properties of $N=9$ qubits Heisenberg $XX$ and $XXZ$
models under the presence of  a polarized  magnetic field in $xz$ plane by means of concurrence concept.
We perform a systematic analysis for a wide range of the system parameters.
Our results suggest that the global phase regions which separate the entangled and non-entangled regions
sensitively depend upon the spin-spin interaction term of the $z-$ component of two neighboring
spins $J_{z}/J_{x}$, temperature as well as polarized magnetic field components. Thereby, we think that
polarized magnetic field can be used a control parameter to determine the amount of thermal entanglement
between pair of qubits for different temperatures and spin-spin interaction terms. Moreover, it has been
found that the nearest-neighbor pair of qubits does not point out a
re-entrant type entanglement character when one only deals with the nearest-neighbor pair of qubits.
However, as one considers next-nearest neighbor pair of qubits, it is possible to see the evidences of
re-entrant type entanglement behaviors.
\end{abstract}

\end{frontmatter}
\section{Introduction}\label{introduction}
When the nonlocal quantum correlations become important in a many-level system, one may not extract the complete information
about individual sub-levels although the total information about the whole system is known. In such a case, sub-levels become
strongly correlated with each other, and this fact may allow the information between two distant points (such as two qubits separated
by large distances) to be communicated instantly. This phenomenon is called action at a distance, and such a pair of sub-levels
is called entangled. Formerly, Einstein and his co-authors \cite{einstein}, as well as Schr\"{o}dinger \cite{schrodinger1,schrodinger2} argued
that this ``spooky action at a distance'' is a direct consequence of incomplete  nature of quantum mechanics. However,
after three decades, Bell \cite{bell} showed that Einstein's realist idea based on the locality is wrong and it pioneered
consecutive  experimental realizations which proved that the predictions of quantum mechanics are true regarding the
entanglement phenomenon as a nonlocal property of nature.

During the last two decades, a great many experimental efforts have been devoted to entanglement phenomenon in a wide
variety of physical systems including entanglement of many photons, mesoscopic systems, and so on \cite{horodecki1}.
Hensen et al. \cite{hensen} very recently produced 245 entangled pairs of electrons (which were 1.3 kilometers
apart from each other) in nine days. They reported that their results rule out large classes of local realist theories.
On the other hand, in the theoretical ground, it was quite a challenge to measure the amount of entanglement between two
correlated sub-systems \cite{horodecki2,peres,hill,gisin,terhal}. The two distinct measures to distinguish between entangled
and separable states are concurrence \cite{wootters} and negativity \cite{vidal}. One should notice that concurrence
cannot be used as a criterion for separability condition for the systems with dimensions larger than $2\otimes3$ in
Hilbert space. Using concurrence as a measure of entanglement between two qubits, models based on localized spins in the
form of Ising, XY and isotropic, as well as anisotropic Heisenberg systems have been widely investigated in the
literature \cite{gunlycke,kamta,sun,asoudeh,zhang,hu,kheirandish,li,akyuz}. In order to observe entanglement phenomenon
in such systems, selected Hamiltonian should include either off-diagonal terms such as anisotropic exchange coupling
and Dzyaloshinskii-Moriya (DM) interaction, and/or inhomogeneous external magnetic fields along the Ising axis.

Apart from these, pairwise entanglement in the systems with three or more qubits \cite{abliz, yang, arnesen,connor,wang5,asoudeh2,wang2,xi,sahintas,glaser,gu,yeo,zhou,cao,wang7,eryigit,min,zhe,qiang,zhao,jia,min2,wang6}
have also been studied in the forms of XX, XY and Heisenberg models, as well as in the forms of their varieties.
According to these works, some important findings can be listed as follows: Under certain conditions, next-nearest-neighbor
entanglement may be larger than nearest-neighbor entanglement near zero temperature \cite{yang}. As the number of
qubits becomes larger than a larger value of external homogeneous magnetic field is needed to observe entanglement. However,
entanglement disappears shortly after the field exceeds some critical value \cite{arnesen}. Moreover, isotropic
Heisenberg spin chain exhibits non-zero concurrence only when the exchange coupling is of antiferromagnetic type \cite{connor,wang5}
whereas if one applies  a magnetic field then the SU(2) symmetry is broken and it becomes possible for a ferromagnetic
isotropic Heisenberg chain to have thermal and ground states which are completely entangled \cite{asoudeh2}. For XX qubit rings
with periodic boundary conditions, Ref.\cite{wang2} also realized that pairwise entanglement between the nearest-neighbor
qubits is invariant under the magnetic field reversal $B\rightarrow-B$, and that for the same model containing ``even number
of qubits'', bipartite thermal entanglement between neighboring sites should be independent of both the sign of magnetic fields
and exchange constants. Ref. \cite{wang6} showed for isotropic Heisenberg model that the ground state entanglement becomes
enhanced (diminished) with increasing number of qubits in odd (even)-numbered qubit rings. It is also possible to distinguish
between thermally entangled and separable states via examining macroscopic properties such as specific heat and magnetic
susceptibility which can play the role of some kind of entanglement witness \cite{vertesi}.

There are also some other works dealing with entanglement properties of qubit-qutrit and qutrit-qutrit chains \cite{osenda,sun2,wang3,zhang2,sun3,guo,yan,akyuz2,ma,albayrak,guo2,abgaryan,carrillo}. In an extended work, Wang et
al. \cite{wang4} studied the entanglement in a spin-1/2 and spin-$s$ ferrimagnetic chain in which they reported that as the
magnitude of spin-$s$ increases then the temperature value at which entanglement vanishes becomes higher whereas the
ground state thermal entanglement for small-$s$ chains is enhanced. Similarly, Ref. \cite{su} showed that threshold
temperature at which entanglement vanishes increases with increasing spin magnitude.

In practice, it is a difficult task to control the entanglement in a system by manipulating the exchange interactions,
and in some cases, the control of the magnitude and direction of externally applied magnetic field
proved useful for tuning the entanglement of a spin chain \cite{terzis,wang,ling,huang}. Therefore, in the present paper,
our aim is to clarify the entanglement phenomena in $N-$ qubit $XX$ and $XXZ$ chains in the presence of magnetic fields applied in both
longitudinal (i.e. easy axis) and transverse (hard axis) directions. The outline of the paper can be summarized as
follows: In Sec. \ref{formulation} we define our model. Numerical results are presented in Sec. \ref{discussion}.
Finally Sec. \ref{conclusion} contains our conclusions.

\section{Formulation}\label{formulation}
We consider 1D Heisenberg $XX$ and $XXZ$ spin chain systems consisting of $N-$ spin-1/2 particles
interacting with nearest neighbor interaction. Each qubit in the system is under the influence of a
polarized magnetic field applied in $xz$ plane. Within the open boundary condition, the Hamiltonian
of a such system can be  described as follows:
\begin{equation}\label{eq1}
\mathcal{H}=\sum_{i=1}^{N-1}\left[ J_{x}\left(\sigma_{i}^{x}\sigma_{i+1}^{x}+\sigma_{i}^{y}\sigma_{i+1}^{y}\right)+
J_{z}\sigma_{i}^{z}\sigma_{i+1}^{z}\right]+h_{x}\sum_{i=1}^{N}\sigma_{i}^{x}+h_{z}\sum_{i=1}^{N}\sigma_{i}^{z}
\end{equation}
where $\sigma_{i}^{x}, \sigma_{i}^{y}$ and $\sigma_{i}^{z}$ are the Pauli spin operators at the site $i$.
$J_{x}$ and $J_{z}$ are  spin-spin exchange interaction terms of the $x-$ and $z-$ components of two neighboring
spins which are selected to be as $J_{x}>0$ and $J_{z}>0$, i.e., antiferromagnetic type interaction.
Throughout the work, we have fixed the total number of particle $N=9$ and value of spin-spin
interaction $J_{x}=1$. $h_{x}=h_{0}\cos\theta$  and $h_{z}=h_{0}\sin\theta$ are the components of the
external polarized field. Here, $h_{0}$ and $\theta$ are the amplitude of the
field and the angle between the field and its corresponding component of the Pauli
spin operators, respectively. In Eq. (\ref{eq1}), the first summation is over the
nearest neighbor qubits while the second and third ones are over all of the qubits.
The Hamiltonian considered here covers $XX$ model with $J_{z}=0$ and isotropic Heisenberg model with $J_{z}=1$ in
the limit cases. There are several studies in the literature regarding the limit cases and
some variants of the model Hamiltonian \cite{connor, wang5, asoudeh2, wang2,  xi, glaser, cao, wang7,
eryigit,   min,zhao, min2, su, Sheng,   Stauber, Canosa1, Canosa2, Alcaraz}.

The physical properties of the system in thermal equilibrium can be obtained via the density matrix operator,
which can be defined as follows:
\begin{equation}\label{eq2}
\rho=\frac{1}{Z}\exp(-\beta\mathcal{H})
\end{equation}
here $\beta=1/k_{B}T$, $T$ and $k_{B}$ are the temperature and Boltzmann constant, respectively. $Z$ is the partition function
which is defined by the trace of the density matrix as $Z=Tr\exp\left({-\beta \mathcal{H}}\right)$. The basic strategy for
obtaining the concurrence (as a measure of bipartite entanglement) for two qubits (say $\sigma_m$ and $\sigma_n$) is, constructing
the reduced density matrix by tracing out the other spins, and calculating  $C_{mn}$ via \cite{wootters}
\begin{equation}\label{eq3}
C_{mn}(R)=max\left\{2\lambda_{1}-\sum_{i=1}^{4}\lambda_{i},0\right\}
\end{equation}
where $\lambda_{i}'$s are the square roots of the eigenvalues of the  operator $R$ which is obtained by
\begin{equation}\label{eq4}
R=\rho^{(mn)}\left( \sigma_{m}^{y}\times  \sigma_{n}^{y} \right)\rho^{(mn)\ast}\left( \sigma_{m}^{y}\times \sigma_{n}^{y} \right)
\end{equation}
in descending order.  Here, $\sigma_{m}^y$ is the $y-$ component of the Pauli spin matrix related to the spin $m$, $\rho^{mn}$
is reduced density matrix, and $\ast$ denotes the complex conjugation. The value of concurrence  varies from $C=0$ for
a separable state (no entanglement) to $C=1$ for a  maximally entangled state.

\section{Results and Discussion}\label{discussion}
\begin{figure}[!h]
\begin{center}
\includegraphics[width=4.cm]{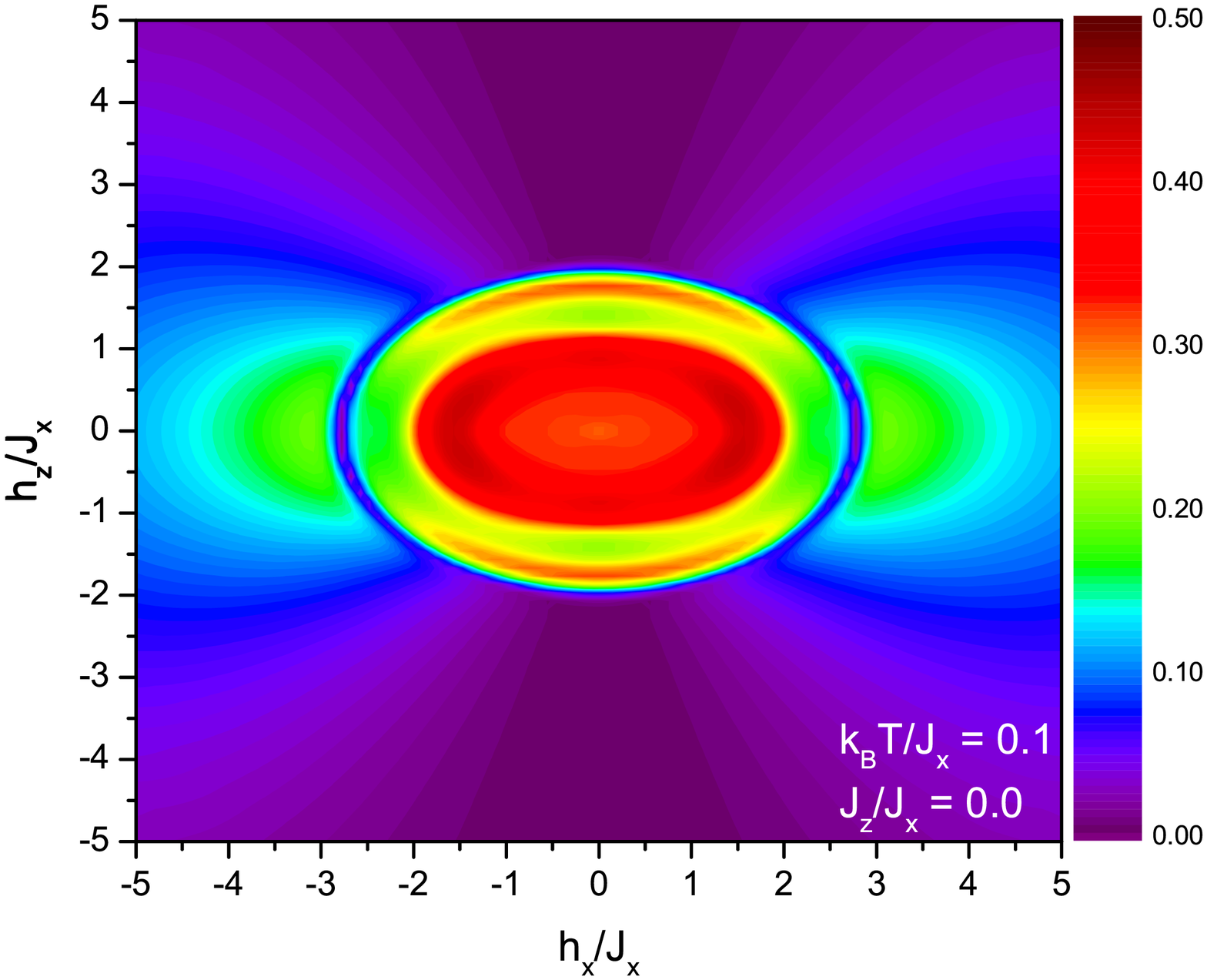}
\includegraphics[width=4.cm]{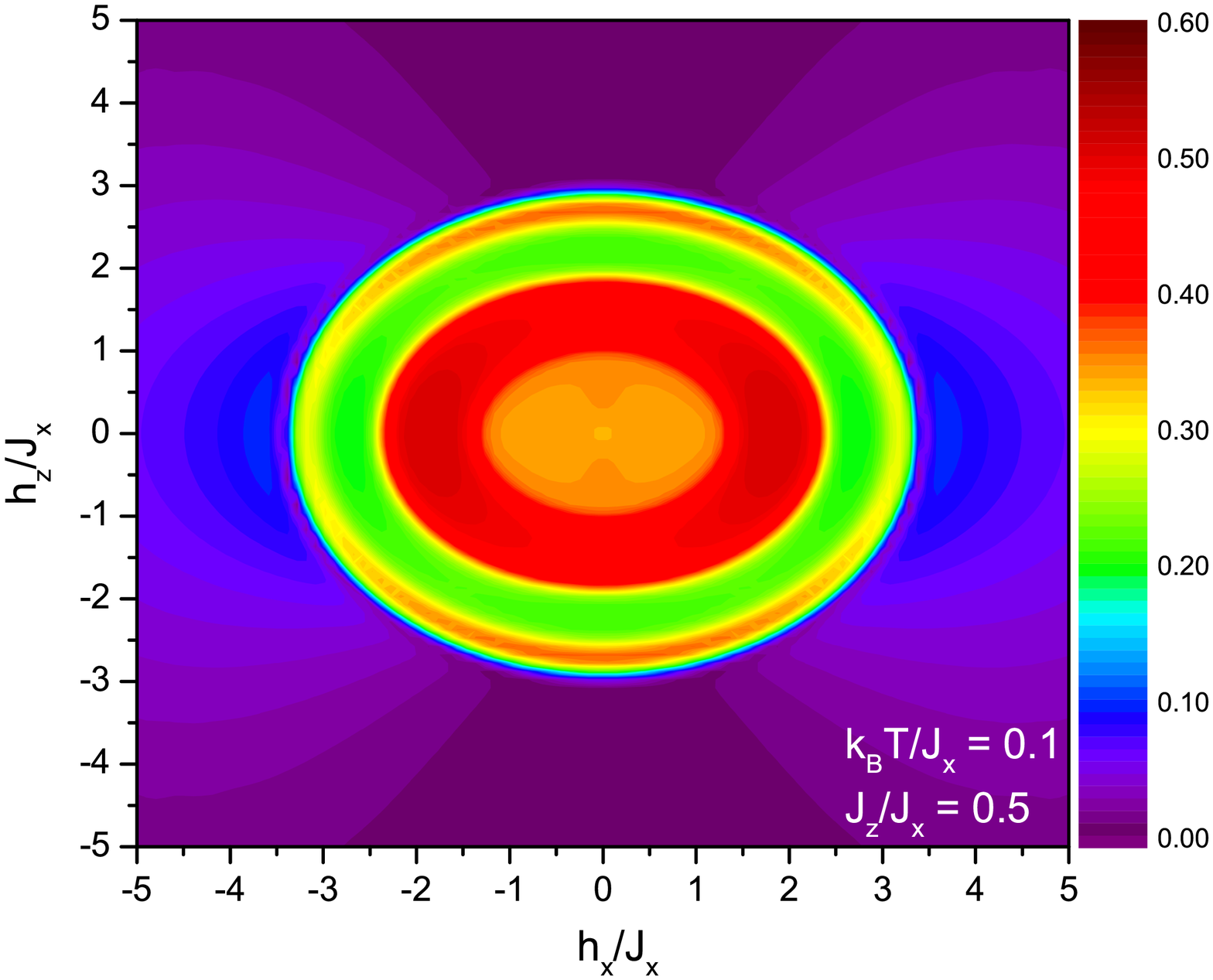}
\includegraphics[width=4.cm]{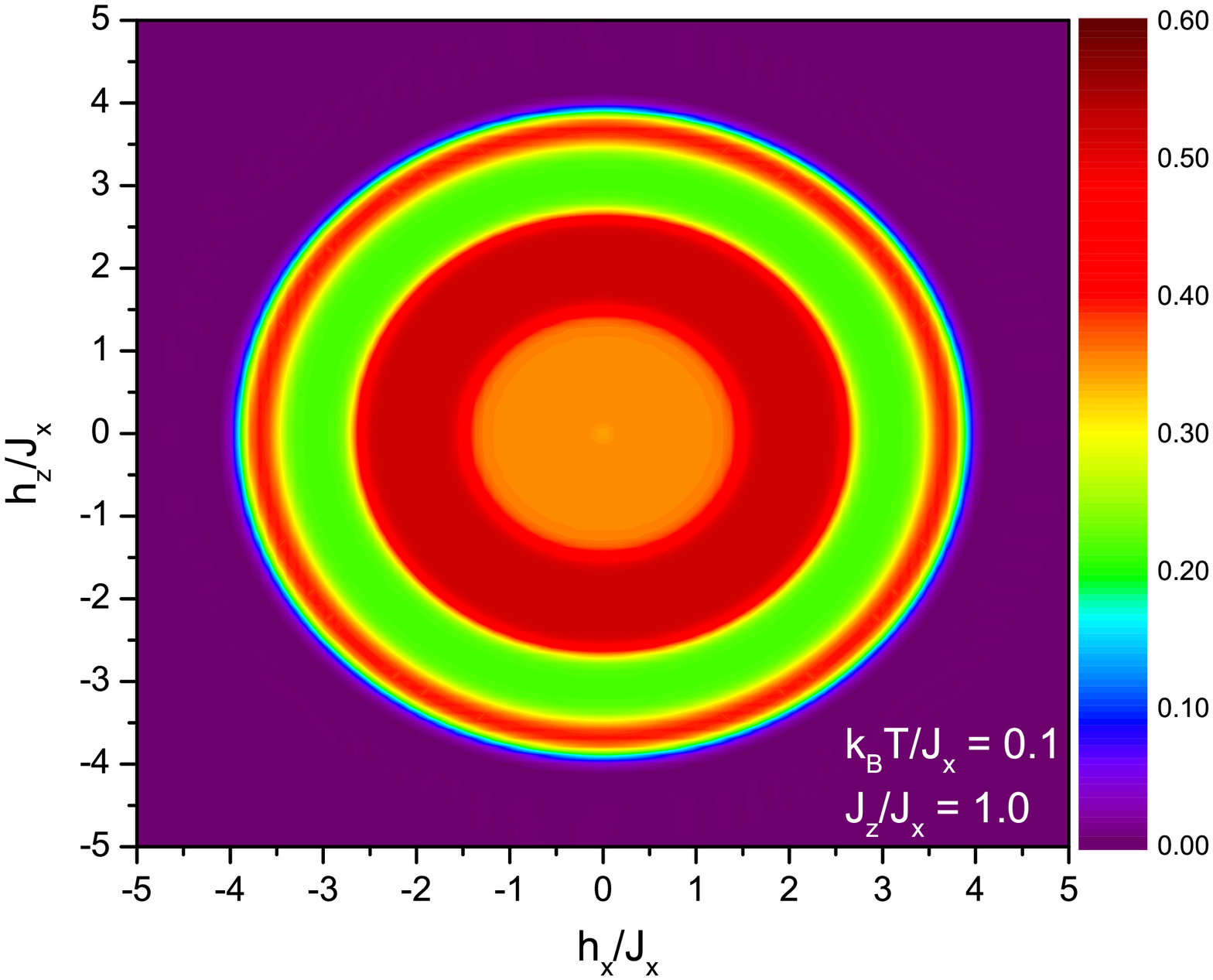}
\includegraphics[width=4.cm]{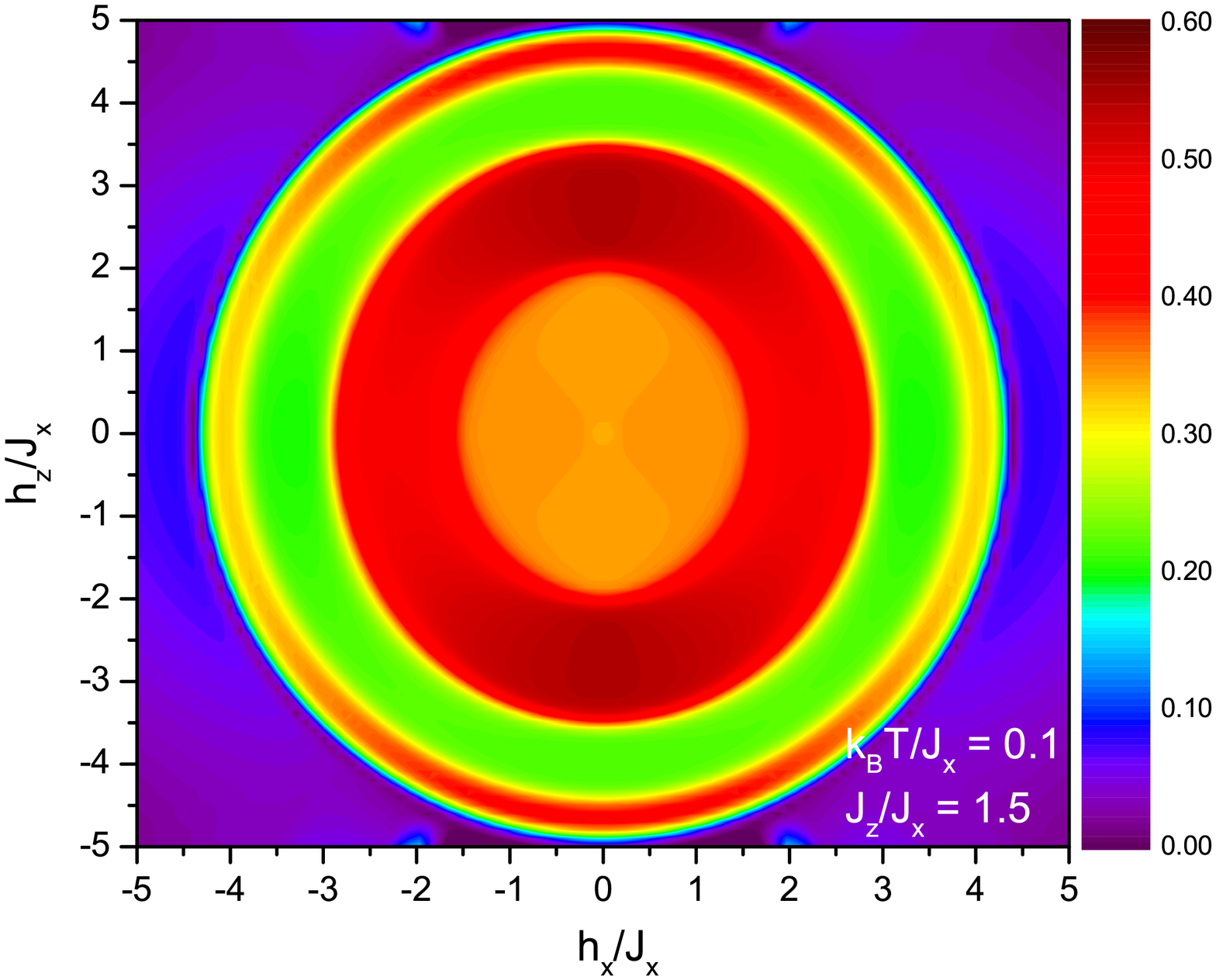}
\caption{Contour maps of magnetic field dependencies of the nearest-neighbor concurrence $C_{56}$
for varying values of the $J_{z}/J_{x}$ ratio such as (a) $J_{z}/J_{x}=0.0$,
(b) 0.5, (c) 1.0 and (d) 1.5, respectively. The figures are plotted for the value
of $k_{B}T/J_{x}=0.1$.}\label{Fig1}
\end{center}
\end{figure}

In this section, we will focus our attention on the thermal entanglement properties of the $N=9$ qubits
$XX$ and $XXZ$ Heisenberg spin chains under the existence of a polarized magnetic field.
We especially study the pairwise entanglement between central qubit and its nearest neighbor entanglement
$C_{56}$ and as well as next-nearest entanglement $C_{57}$ to show how the applied magnetic fields, interaction
constant of the $z-$ component and the temperature affect the natures of entanglement features of the system.
For this aim, the boundaries in related planes separating entangled and non-entangled
regions will be given for the selected Hamiltonian parameters. The physical mechanisms
underlying of these types of observations will be discussed in detail. Depending on the considered Hamiltonian parameters,
in the entangled region, the concurrence has a non-zero value corresponding to the presence of a strongly correlated two qubit pair.
Contrary to this, concurrence vanishes in the non-entangled region. Keeping these facts in mind,
we illustrate the boundaries in the $(h_{x}/J_{x}-h_{z}/J_{x})$ contour map in Figs. 1(a-d) with varying
$J_{z}/J_{x}$ parameters such as $J_{z}/J_{x}=0.0$ (a), 0.5 (b), 1.0 (c) and  1.5 (d), respectively.
The figures  are presented for the nearest-neighbor concurrence $C_{56}$, and for
the relatively low value of temperature, i.e. for  $k_{B}T/J_{x}=0.1$. It is clear from figure \ref{Fig1}(a) that
along the $h_{x}/J_{x}=0.0$  axis, when the value of $|h_{z}/J_{x}|$ increases
starting from zero, the concurrence $C_{56}$ tends to rise until it reaches a certain value for a given
temperature. Then, $C_{56}$ begins to decrease with further increment in value of $|h_{z}/J_{x}|$.
Moreover, as the value of $|h_{z}/J_{x}|$ is sufficiently large, the entanglement between qubits $5$ and $6$
disappears, hence, $C_{56}$ vanishes. These behaviors appear symmetrically at two sides
of $h_{z}/J_{x}$. Along the $h_{z}/J_{x}=0.0$ axis, magnetic field dependency $(h_{x}/J_{x})$ of $C_{56}$
exhibits a  nearly similar character to the previous one mentioned above.  The only difference is that the
entanglement between  qubits $5$ and $6$ still lives as the value  of $|h_{x}/J_{x}|$ is sufficiently too large.
Figures \ref{Fig1}(b-d) show effect of the interaction  constant $J_{z}/J_{x}$  on the thermal
entanglement character of  $C_{56}$.

\noindent Based on our numerical results, it has been found that $J_{z}/J_{x}$ parameter plays an important role in
enhancing the entanglement. As the value of $J_{z}/J_{x}$ is too large, much more energy originating from
magnetic fields is needed to drive the system from entangled state to non-entangled one. The aforementioned situations can
be easily seen by comparing the figures \ref{Fig1}(a-d) with each other. The findings also demonstrate that
the geometrical shape where the entanglement character exists sensitively depends on the selected $J_{z}/J_{x}$ parameter.
For instance, figures \ref{Fig1}(a) and (b) present an elliptical shape whereas the figure (c) shows a
circular shape entanglement character. However, as depicted in figure \ref{Fig1}(d), the general trend
of entanglement  character again becomes elliptical with further increment in value of $J_{z}/J_{x}$.

\begin{figure}[!h]
\begin{center}
\includegraphics[width=4.cm]{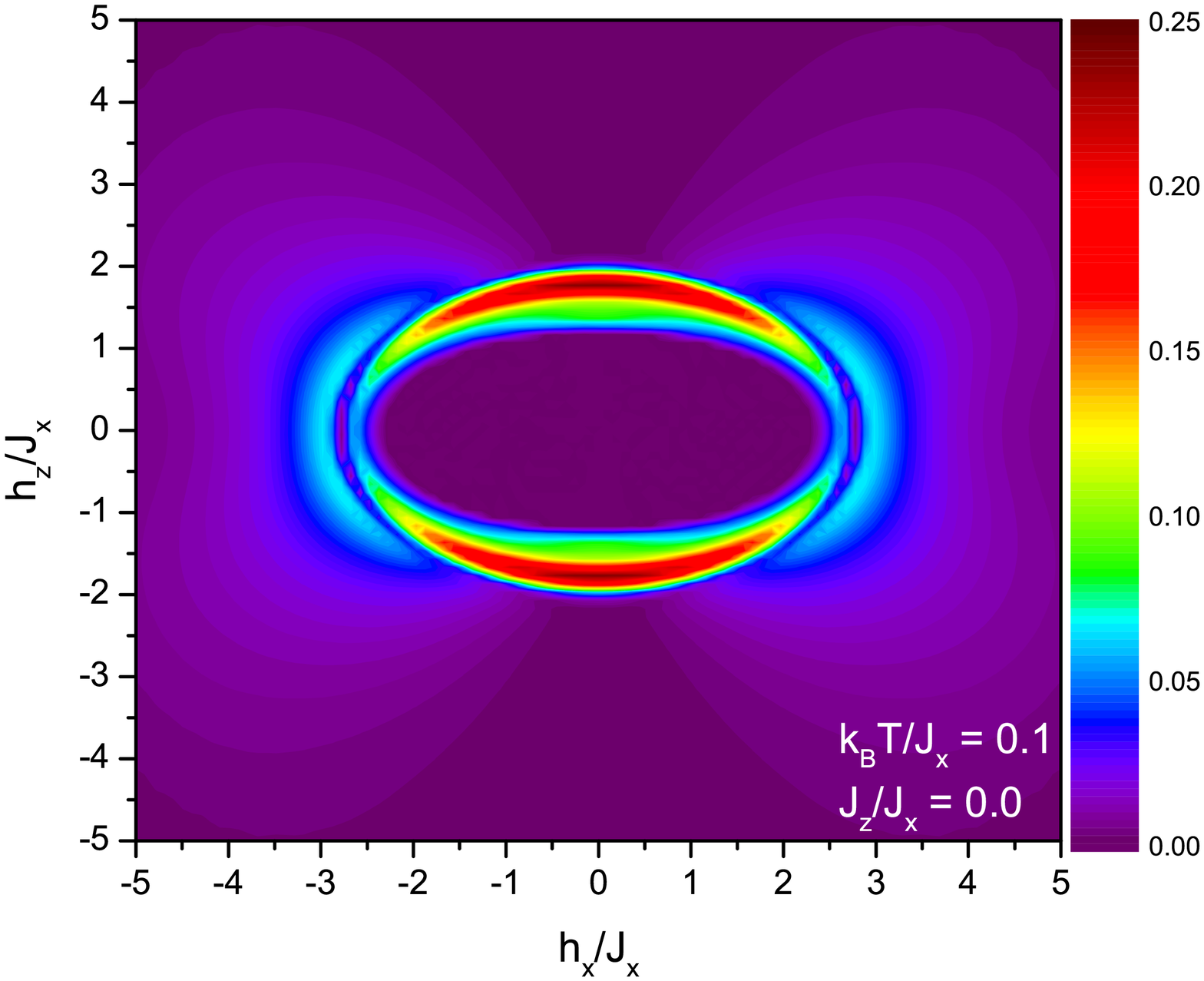}
\includegraphics[width=4.cm]{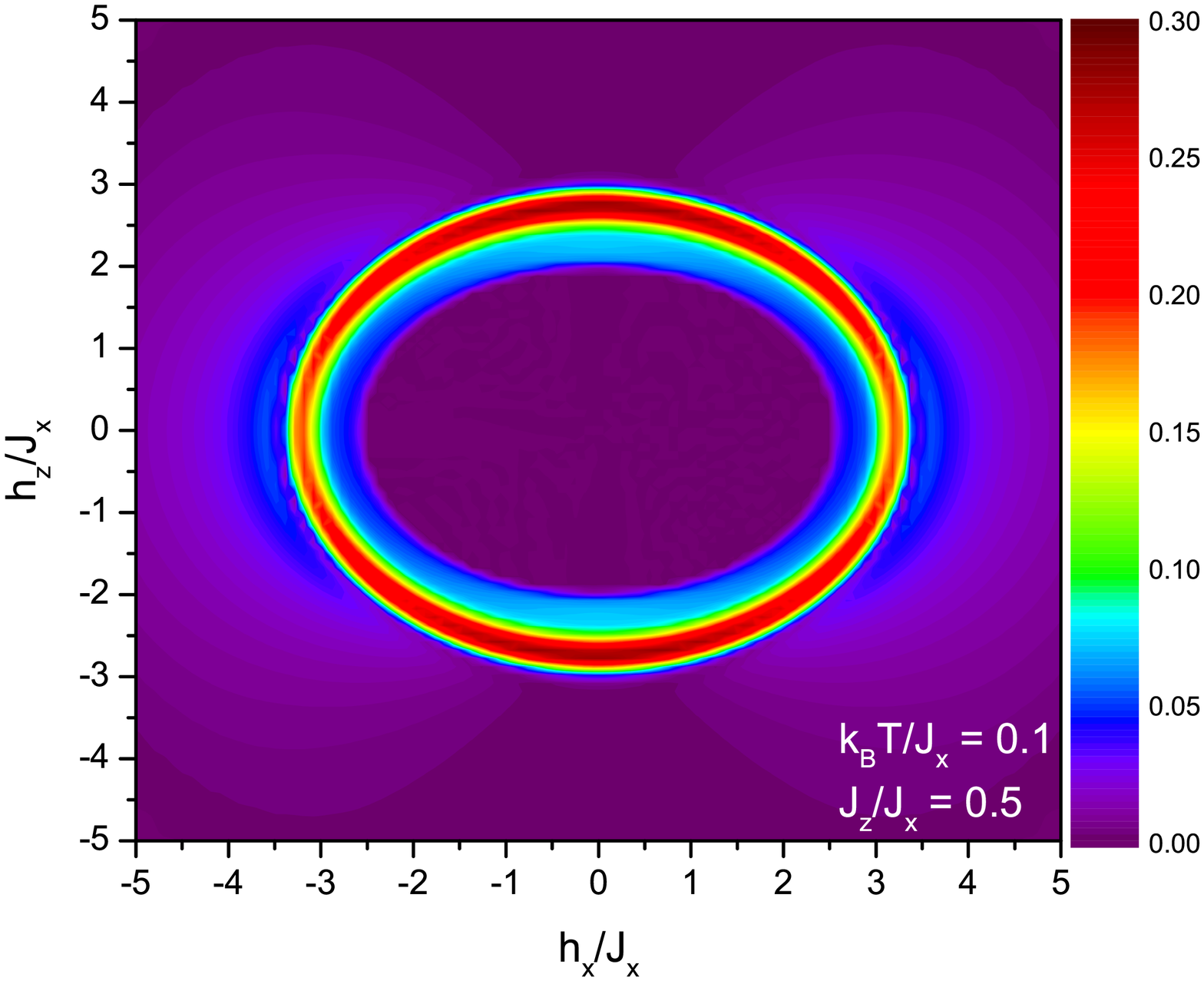}
\includegraphics[width=4.cm]{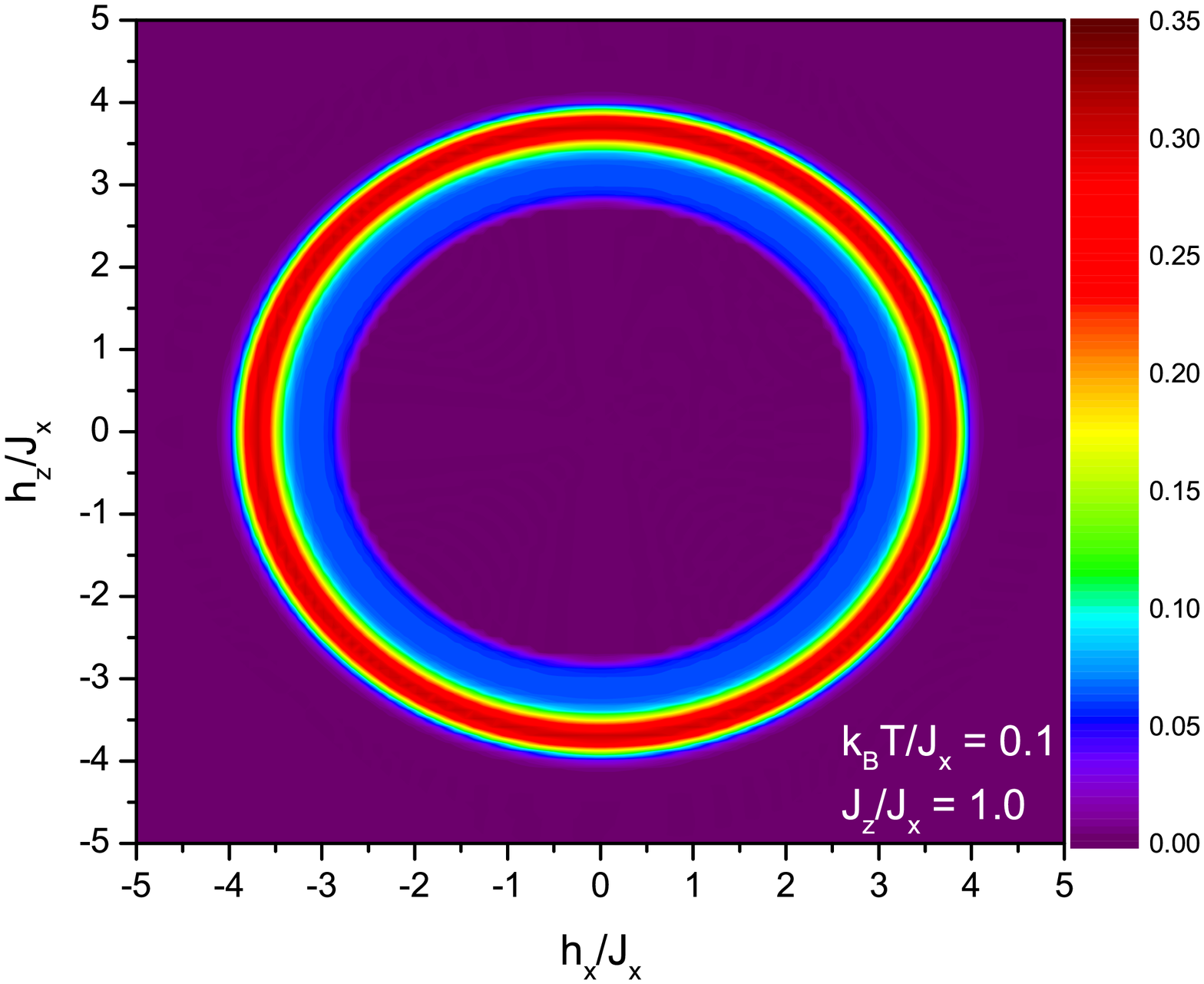}
\includegraphics[width=4.cm]{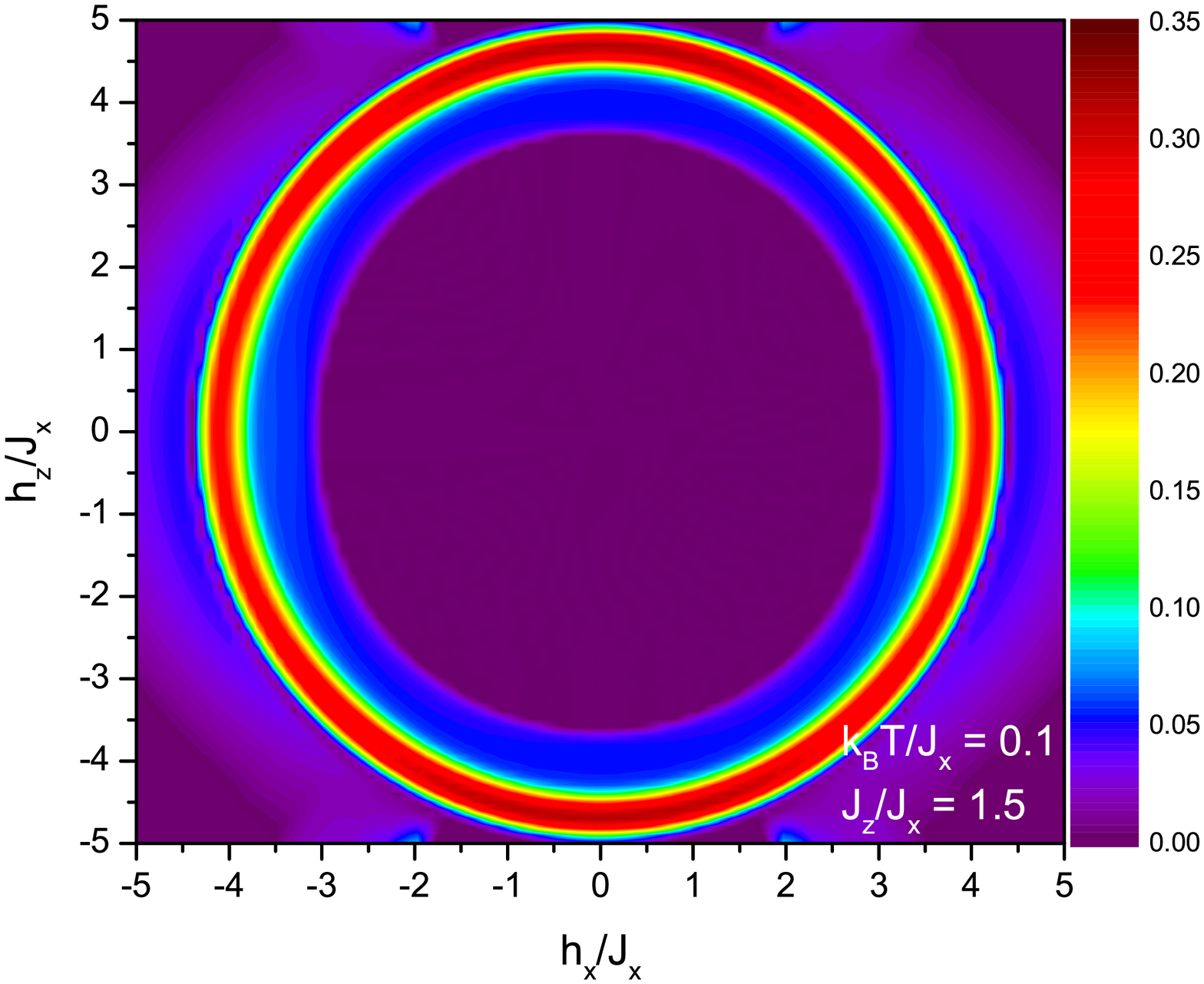}
\caption{Contour maps of magnetic field dependencies of the next nearest-neighbor concurrence $C_{57}$
for the same Hamiltonian parameters with figure (\ref{Fig1}).}\label{Fig2}
\end{center}
\end{figure}

In figure \ref{Fig2}(a-d), we  examine thermal entanglement features of the next-nearest neighbor
concurrence $C_{57}$ of nine-qubit system for the same Hamiltonian parameters with figure (\ref{Fig1}).
In contrary to the figure \ref{Fig1}, in the absence of polarized magnetic field, there is no entanglement
behavior between  qubits $5$ and $7$. However, increment in strength of the magnetic field gives rise to
the existence of an entanglement character. Next, as the strength of applied magnetic field is increased further,
$5$ and $7$ qubits become non-entangled. The mechanism briefly mentioned here is often called the re-entrant type
entanglement behavior. We should indicate that there are several studies in the literature showing a re-entrant
type entanglement character \cite{hu, abliz, glaser}.

\begin{figure}[!h]
\begin{center}
\includegraphics[width=4.cm]{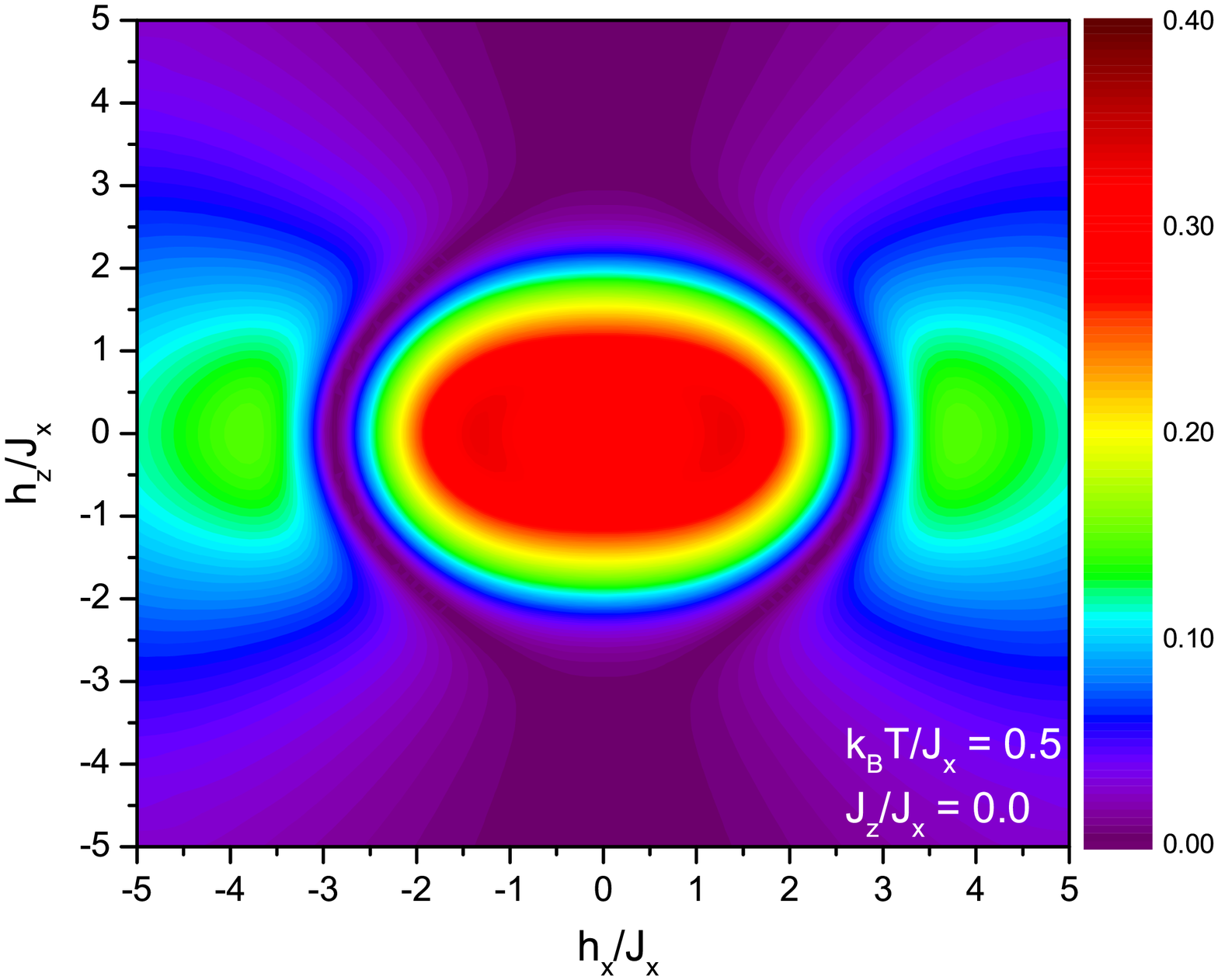}
\includegraphics[width=4.cm]{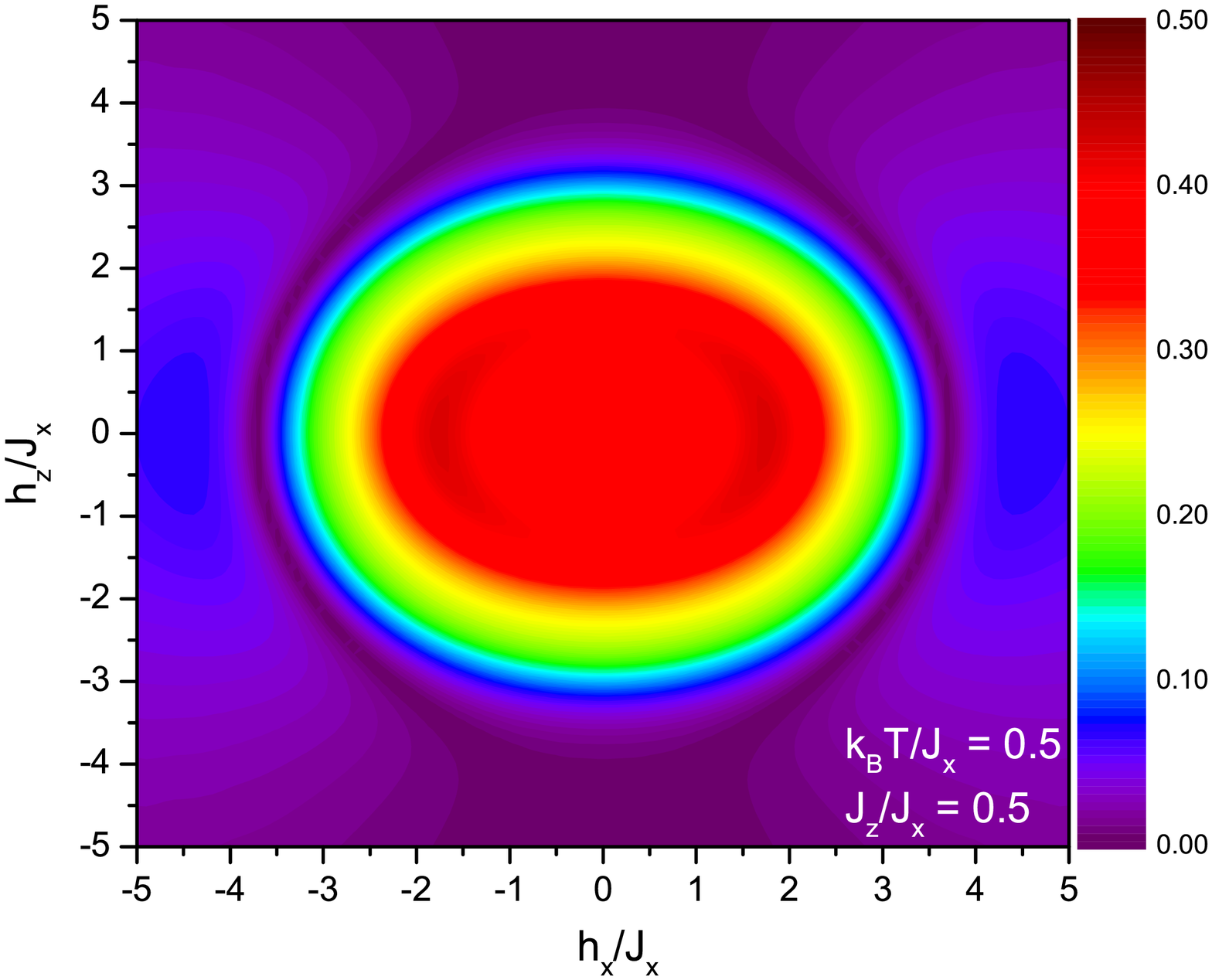}
\includegraphics[width=4.cm]{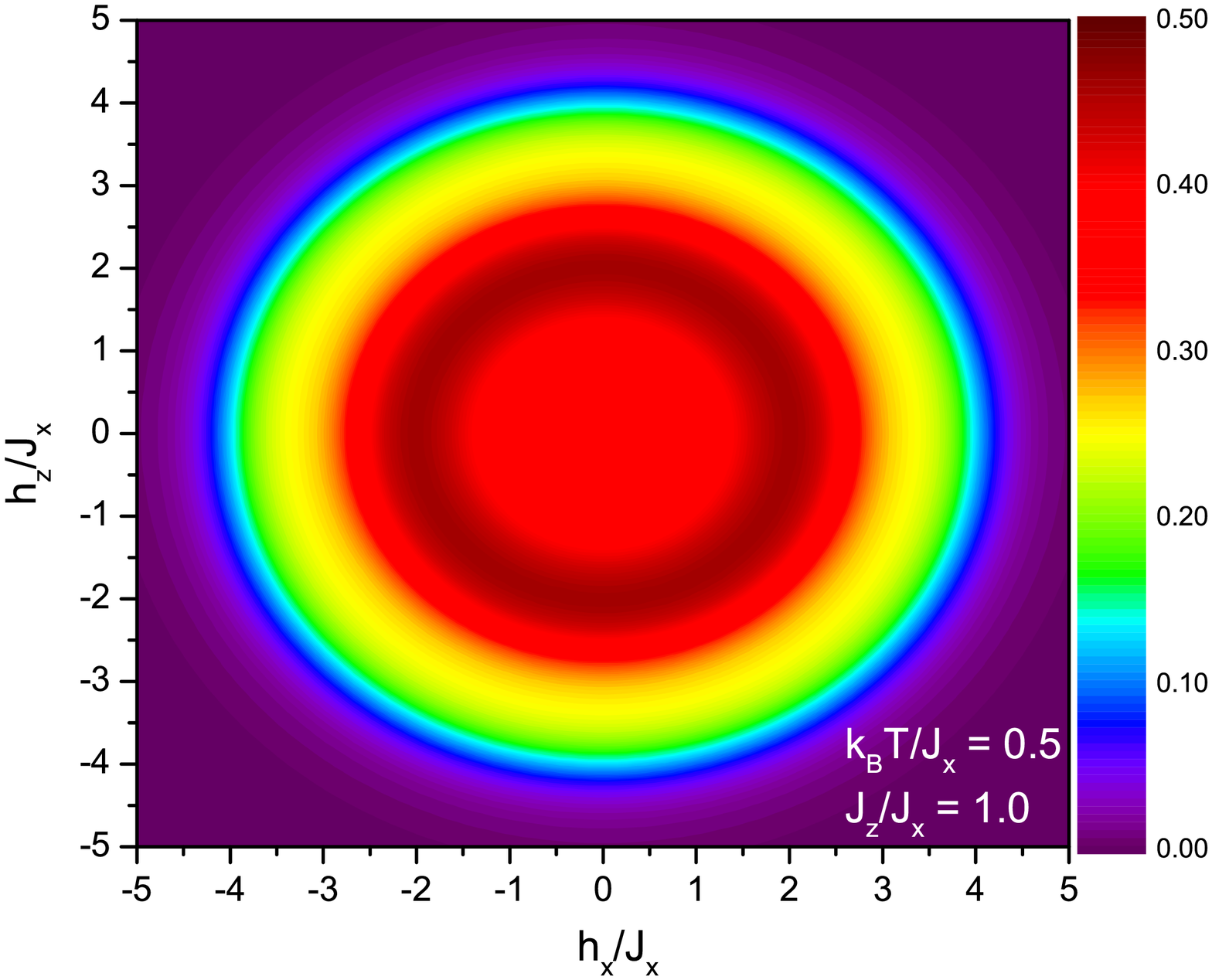}
\includegraphics[width=4.cm]{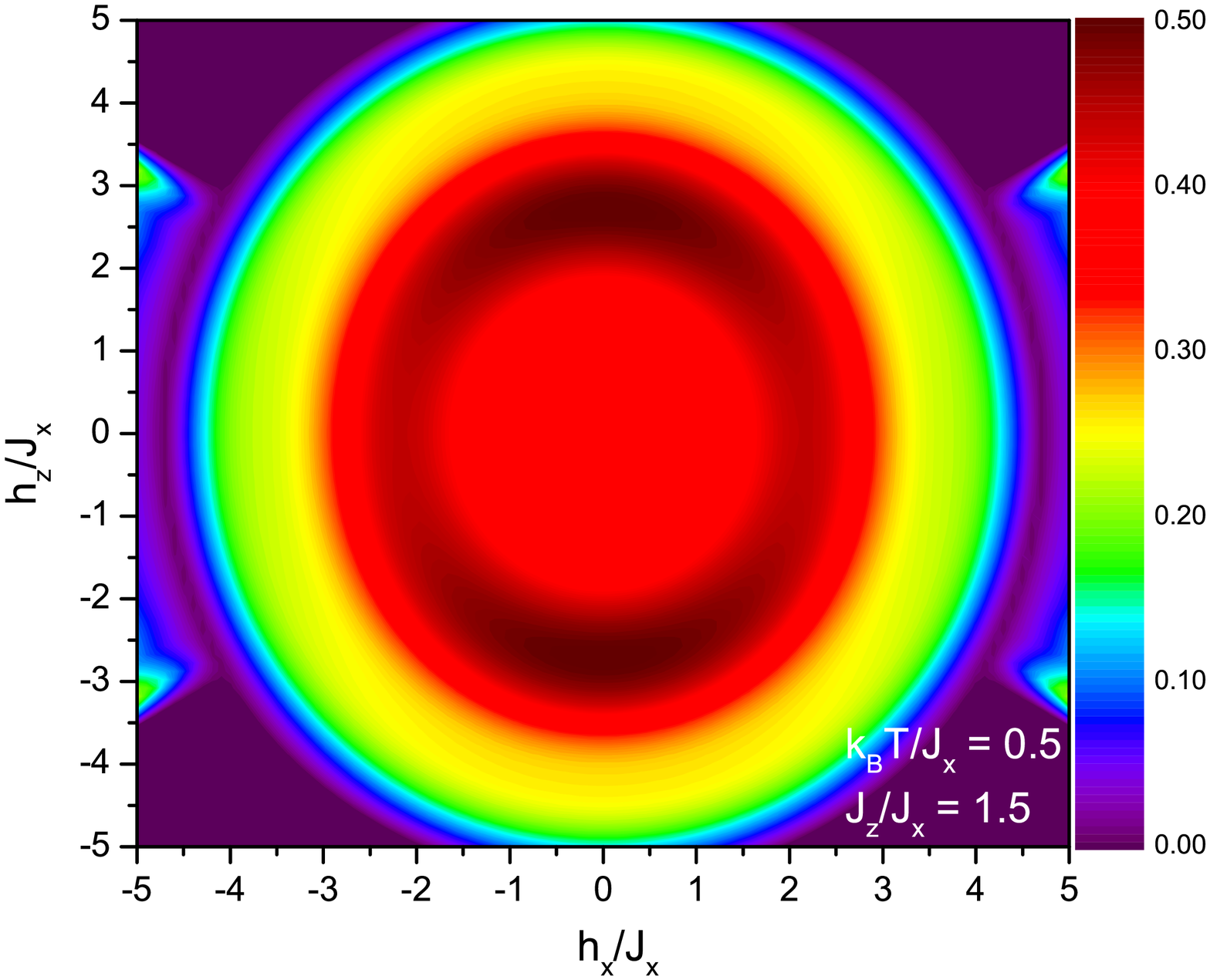}
\caption{Contour maps of magnetic field dependencies of the nearest-neighbor concurrence $C_{56}$
for varying values of the $J_{z}/J_{x}$ ratio such as (a) $J_{z}/J_{x}=0.0$,
(b) 0.5, (c) 1.0 and (d) 1.5, respectively. The figures are plotted for the value
of $k_{B}T/J_{x}=0.5$.}\label{Fig3}
\end{center}
\end{figure}

\begin{figure}[!h]
\begin{center}
\includegraphics[width=4.cm]{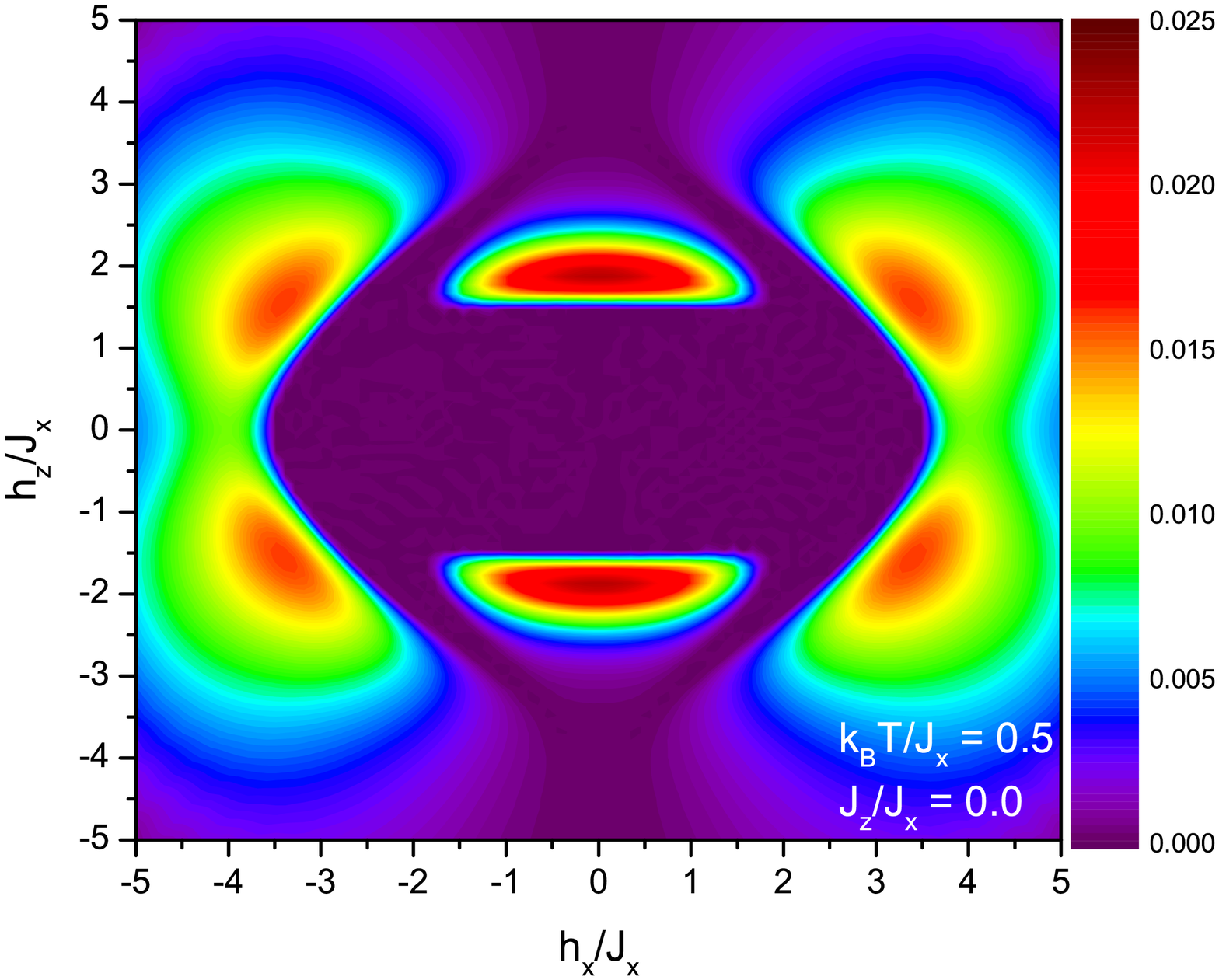}
\includegraphics[width=4.cm]{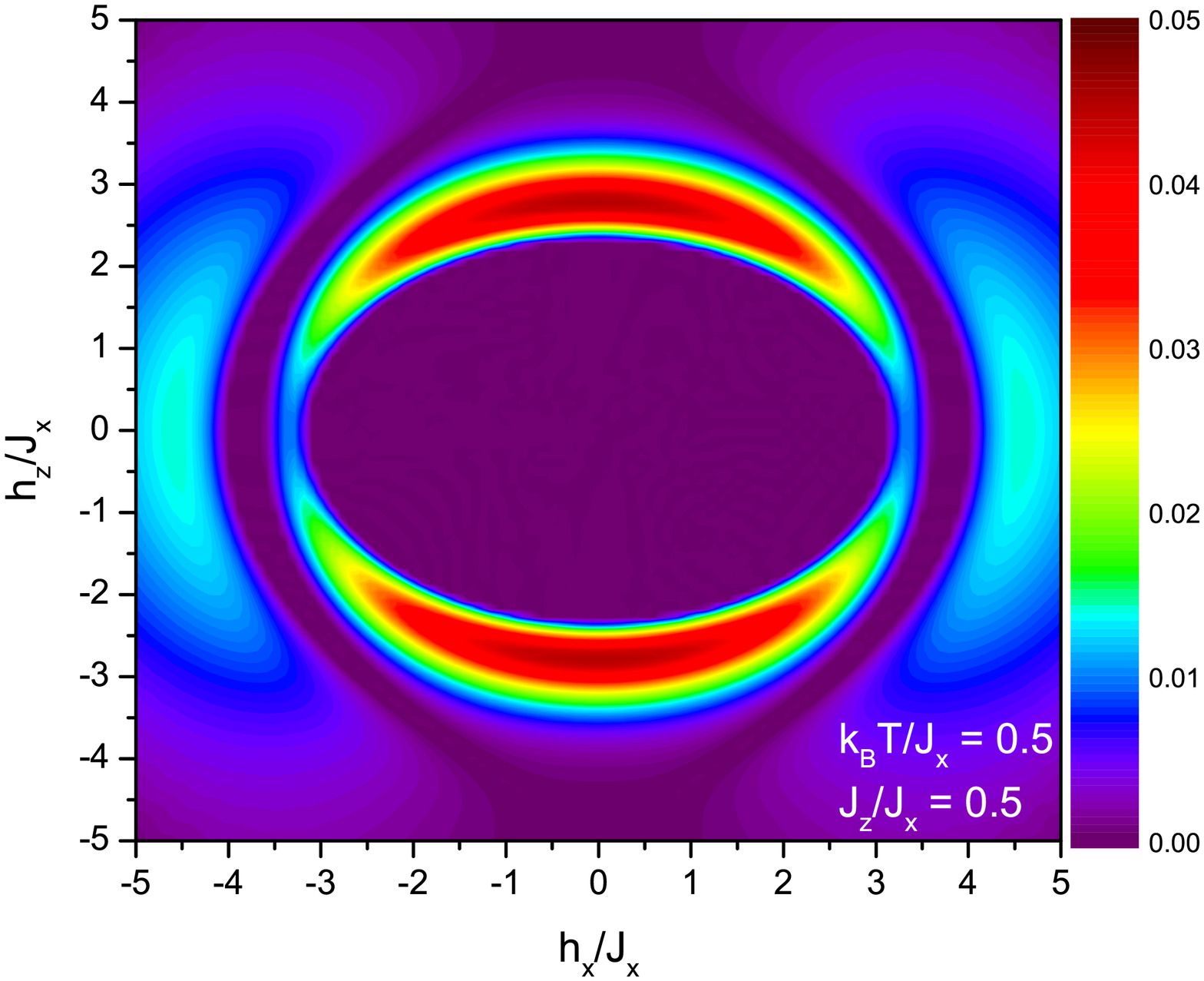}
\includegraphics[width=4.cm]{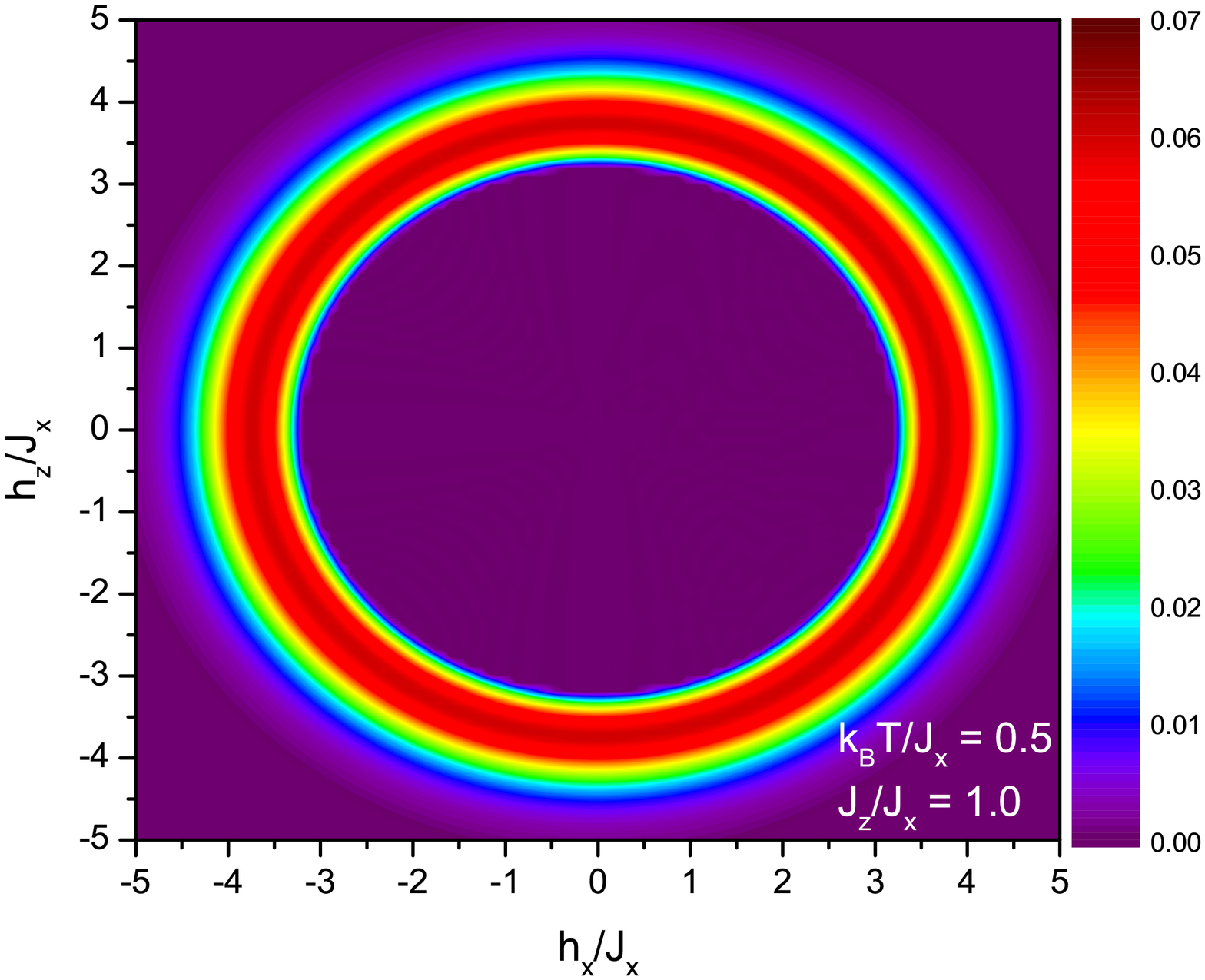}
\includegraphics[width=4.cm]{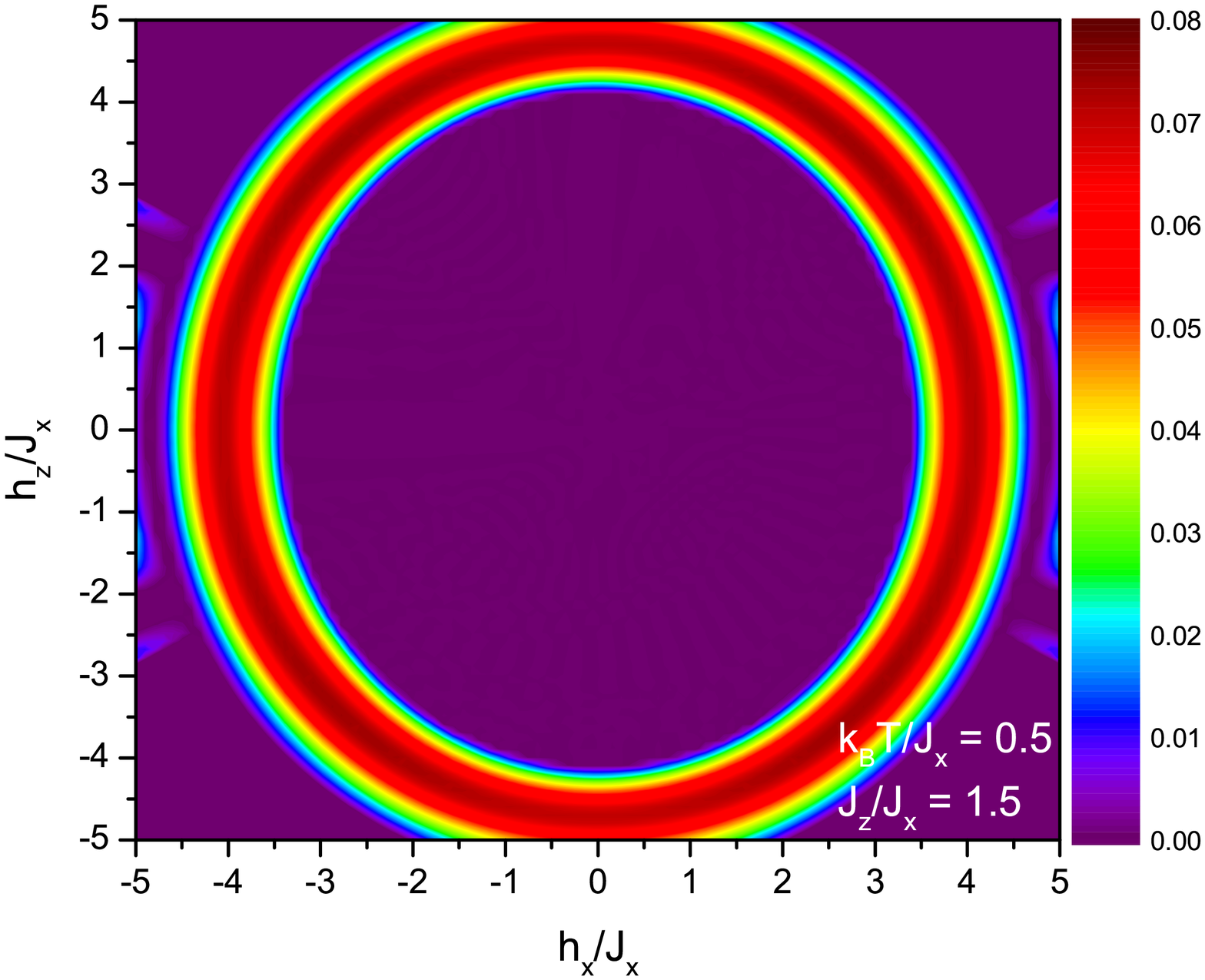}
\caption{Contour maps of magnetic field dependencies of the next nearest-neighbor concurrence $C_{57}$
for the same Hamiltonian parameters with figure (\ref{Fig3}).}\label{Fig4}
\end{center}
\end{figure}

In order to elucidate the temperature influences on the entanglement properties of the studied system,
we plot figures (\ref{Fig3}) and (\ref{Fig4}) corresponding to the  concurrences $C_{56}$
and $C_{57}$, respectively. The figures are demonstrated for the value of $k_{B}T/J_{x}=0.5$.
The raising temperature  shows a tendency to reduce and destruct the quantum
correlations between the considered qubits for the fixed sets of Hamiltonian parameters. If one compares the
figures (\ref{Fig1}) and (\ref{Fig3}) with each other,  it can be easily seen that the concurrence $C_{56}$
prominently decreases  when the value of temperature is increased because the thermal energy is dominant
against the spin-spin interactions. We should also note that although an increment in the value of temperature
causes a quantitative change in concurrence $C_{56}$, it does not lead to a
change in characteristic behavior. It is beneficial to notice that similar type observations originating from the thermal
agitations have been reported in Ref. \cite{gunlycke} where ground state and finite temperature features of
two qubit system under the influence of a polarized magnetic field have been discussed in detail.
Furthermore, by comparing the figures (\ref{Fig2}) and (\ref{Fig4}) with each other,  it is possible to say
that when the temperature increases from $k_{B}T/J_{x}=0.1$ to $0.5$, unusual and interesting thermal entanglement
behaviors occur in next-nearest neighbor concurrence  $C_{57}$. These  dramatic
changes have emerged at the relatively low values of $J_{z}/J_{x}$ such as for $J_{z}/J_{x}=0.0$
and $0.5$. For example, non-entangled region expands in $h_{x}/J_{x}$ and $h_{z}/J_{x}$ plane as well as
a re-entrant type entanglement character takes place along the $h_{x}/J_{x}=0$ lines. It means that the
considered qubits are dragged from non-entangled region to entangled region with increasing value of magnetic field $h_{z}/J_{x}$. Then,
as the energy arising from the magnetic field is increased further, the system displays opposite
behavior than the previous one discussed above.

\section{Conclusion}\label{conclusion}
In conclusion, we have investigated the thermal entanglement properties of $N=9$ qubits
$XX$ and $XXZ$ Heisenberg spin chains within the open boundary condition by making use of concurrence
concept. We design the system such that each qubit in the chain is exposed to the polarized magnetic
field in $xz$ plane. According to our detailed investigation, the most conspicuous findings mentioned in the
present study can be listed as follows. The boundaries which separate the entangled and non-entangled regions
sensitively depend upon the spin-spin interaction term of the $z-$ component of two neighboring
spins $J_{z}/J_{x}$, temperature as well as polarized magnetic field components. For example, $J_{z}/J_{x}$
parameter plays a crucial role in improving the  entanglement depending on the other Hamiltonian parameters.
It allows us to control the amount of the entanglement between the selected pair of qubits by varying the
value of $J_{z}/J_{x}$. Moreover, it is possible to realize a magnetically stimulated entanglement
behavior by selecting the suitable system parameters. By changing the value of polarized
magnetic field, one may create or destruct the quantum correlations between the pair of qubits.
It has also been found that the present system presents a re-entrant type thermal entanglement character
depending on the considered system parameters if one considers the next-nearest neighbor pair of qubits.
Finally, we  deal with the finite temperature influences on the entanglement character of system,
and the raising temperature demonstrate a tendency to shrink the quantum correlations
between the considered qubits for the fixed sets of Hamiltonian parameters.

\section{Acknowledgements}
The numerical calculations reported in this paper were performed at T\"{U}B\.{I}TAK ULAKB\.{I}M (Turkish agency), High Performance and
Grid Computing Center (TRUBA Resources).


\begin{thebibliography}{99}
\bibitem{einstein} A. Einstein, B. Podolsky, N. Rosen,  Phys. Rev. 47 (1935) 777.
\bibitem{schrodinger1} E. Schr\"{o}dinger,  Naturwissenschaften 23 (1935) 807.
\bibitem{schrodinger2} E. Schr\"{o}dinger,  Math. Proc. Cambridge 31 (1935) 555.
\bibitem{bell} J.S. Bell,  Physics 1 (1964) 195.
\bibitem{horodecki1} R. Horodecki, P. Horodecki, M. Horodecki, K. Horodecki,
Rev. Mod. Phys. 81 (2009) 865, and the references therein.
\bibitem{hensen}  B. Hensen, H. Bernien, A.E. Dr\'{e}au, A. Reiserer,
N. Kalb, M.S. Blok, J. Ruitenberg, R.F.L. Vermeulen, R.N. Schouten, C. Abell\'{a}n,
W. Amaya, V. Pruneri, M.W. Mitchell, M. Markham, D.J. Twitchen, D. Elkouss, S. Wehner,
T.H. Taminiau, R. Hanson, Nature 526 (2015) 682.
\bibitem{horodecki2} M. Horodecki, P. Horodecki, R. Horodecki, Phys. Lett. A  223 (1996) 1.
\bibitem{peres} A. Peres,  Phys. Rev. Lett. 77 (1996) 1413.
\bibitem{hill} S. Hill, W.K. Wootters,  Phys. Rev. Lett. 78 (1997) 5022.
\bibitem{gisin} N. Gisin, Phys. Lett. A 210 (1996) 151.
\bibitem{terhal} B.M. Terhal,  Phys. Lett. A 271 (2000) 319.
\bibitem{wootters} W.K. Wootters,  Phys. Rev. Lett. 80 (1998) 2245.
\bibitem{vidal} G. Vidal, R.F. Werner,  Phys. Rev. A  65 (2002) 032314.
\bibitem{gunlycke} D. Gunlycke, V.M. Kendon, V. Vedral,  Phys. Rev. A 64 (2001) 042302.
\bibitem{kamta} G.L. Kamta, A.F. Starace,  Phys. Rev. Lett. 88 (2002) 107901.
\bibitem{sun} Y. Sun, Y. Chen, H. Chen,  Phys. Rev. A 68 (2003) 044301.
\bibitem{asoudeh} A. Asoudeh, V. Karimipour, Phys. Rev. A  71 (2005) 022308.
\bibitem{zhang} G.F. Zhang, S.S. Li, Phys. Rev. A 72 (2005) 034302.
\bibitem{hu} Z.N. Hu, S.H. Youn, K. Kang, C.S. Kim, J. Phys. A: Math. Gen. 39 (2006) 10523.
\bibitem{kheirandish} F. Kheirandish, S.J. Akhtarshenas, H. Mohammadi,  Phys. Rev. A 77 (2008) 042309.
\bibitem{li} D.C. Li, X.P. Wang, Z.L. Cao, J. Phys.: Condens. Matter 20 (2008) 325229.
\bibitem{akyuz} C. Aky\"{u}z, E. Aydiner,  Chin. Phys. Lett.  25 (2008) 1557.
\bibitem{abliz} A. Abliz, J.T. Cai, G.F. Zhang, G.S. Jin,  J. Phys. B: At. Mol. Opt. Phys. 42 (2009) 215503.
\bibitem{yang} G.H. Yang,  L. Zhou,  Commun. Theor. Phys. (Beijing, China) 49 (2008) 1635.
\bibitem{arnesen} M.C. Arnesen, S. Bose, V. Vedral,  Phys. Rev. Lett. 87 (2001) 017901.
\bibitem{connor} K. M. O'Connor, W.K. Wootters, Phys. Rev. A 63 (2001) 052302.
\bibitem{wang5} X. Wang, P. Zanardi, Phys. Lett. A 301 (2002) 1.
\bibitem{asoudeh2} M. Asoudeh, V. Karimipour, Phys. Rev. A 70 (2004) 052307.
\bibitem{wang2} X. Wang,  Phys. Rev. A 66 (2002) 034302.
\bibitem{xi} X.Q. Xi, W.X. Chen, S.R. Hao, R.H. Yue,  Physics Letters A 300 (2002) 567.
\bibitem{sahintas} A. Sahintas, C. Aky\"{u}z,  Physica A 448 (2016) 10.
\bibitem{glaser} U. Glaser, H. B\"{u}tner, H. Fehske, Phys. Rev. A 68 (2003) 032318.
\bibitem{gu} S.J. Gu, H.Q. Lin, Y.Q. Li, Phys. Rev. A 68 (2003) 042330.
\bibitem{yeo} Y. Yeo, Phys. Rev. A 68 (2003) 022316.
\bibitem{zhou} L. Zhou, H.S. Song, Y.Q. Guo, C. Li,  Phys. Rev. A 68 (2003) 024301.
\bibitem{cao} M. Cao, S. Zhu, Phys. Rev. A 71 (2005) 034311.
\bibitem{wang7} X.G. Wang, Physics Letters A 334 (2005) 352.
\bibitem{eryigit} R. Eryigit, Y.G\"{u}nd\"{u}c, R. Eryigit, Physics Letters A 349 (2006) 37.
\bibitem{min} C. Min, Z.S. Qun, Chin. Phys. Lett. 23 (2006) 2888.
\bibitem{zhe} S. Zhe and W.X. Guang, Commun. Theor. Phys. 45 (2006) 61.
\bibitem{qiang} X.X. Qiang, C.W. Xue, L. Qi, Y.R. Hong, Commun. Theor. Phys. 48 (2007) 1009.
\bibitem{zhao} X.Y. Zhao, L. Zhou, Int. J. Theor. Phys. 46 (2007) 2437.
\bibitem{jia} S.C. Jia, C.W. Wen, L.T. Kun, H.Y. Xia, L. Hong, X.Y. Jie, Chin. Phys. B 17 (2008) 1674.
\bibitem{min2} C. Min, L.Y. Sheng and Z.S. Qun, Commun. Theor. Phys. 51 (2009) 811.
\bibitem{wang6} X. Wang, Phys. Rev. A 66 (2002) 044305.
\bibitem{vertesi} T. V\'{e}rtesi, E. Bene, Phys. Rev. B 73 (2006) 134404.
\bibitem{osenda} O. Osenda, G.A. Raggio,  Phys. Rev. A 72 (2005) 064102.
\bibitem{sun2} Z. Sun, X.G. Wang, Y.Q. Li,  New J. Phys. 7 (2005) 83.
\bibitem{wang3} X. Wang, H.B. Li, Z. Sun, Y.Q. Li,  J. Phys. A Math. Gen.  38 (2005) 8703.
\bibitem{zhang2}G.F. Zhang, J.Q. Liang, G.E. Zhang, Q.W. Yan, Eur. J. Phys. D 32 (2005) 409.
\bibitem{sun3} Z. Sun, X.G. Wang, A.Z. Hu, Y.Q. Li,  Physica A  370 (2006) 483.
\bibitem{guo} J.L. Guo, X.L. Huang, H.S. Song,  Phys. Scr. 76 (2007) 327.
\bibitem{yan} Z. Yan, Z. Shi-Qun, H. Xiang, Chin. Phys. 16 (2007) 2229.
\bibitem{akyuz2} C. Aky\"{u}z, E. Aydiner, \"{O}. M\"{u}stecaplioglu, Optics Commun. 281 (2008) 5271.
\bibitem{ma} X.S. Ma,  Optics Commun. 281 (2008) 484.
\bibitem{albayrak} E. Albayrak,  Chin. Phys. B 19 (2010) 090319.
\bibitem{guo2} K.T. Guo, M.C. Liang, H.Y. Xu, C.B. Zhu,  J. Phys. A: Math. Theor 43 (2010) 505301.
\bibitem{abgaryan} V.S. Abgaryan, N.S. Ananikian, L.N. Ananikian, A.N. Kocharian,   Phys. Scr. 83 (2011) 055702.
\bibitem{carrillo} E.S. Carrillo, R. Franco, J.S. Valencia,  Physica A 390 (2011) 2208.
\bibitem{wang4} X. Wang, Z.D. Wang,  Phys. Rev. A 73 (2006) 064302.

\bibitem{su} X.Q. Su, A.M. Wang, Phys. Lett. A 369 (2007) 196.
\bibitem{terzis} A.F. Terzis, E. Paspalakis, Phys. Lett. A 333 (2004) 438.
\bibitem{wang} Y. Wang, J. Cao, Y. Wang, Phys. Lett. A 342 (2005) 375.
\bibitem{ling} C.W. Ling, Y. Dong, G.S. Jian, Chin. Phys. Lett. 25 (2008) 832.
\bibitem{huang} H.L. Huang, Int. J. Theor. Phys. 50 (2011) 70.
\bibitem{Sheng} L. Y.-Sheng, Commun.   Theor. Phys. 48 (2007) 1017.
\bibitem{Stauber} T. Stauber, F. Guinea, Phys. Rev. A 70 (2004) 022313.
\bibitem{Canosa1} N. Canosa, R. Rossignoli, Phys. Rev. A 73 (2006) 022347.
\bibitem{Canosa2} N. Canosa, R. Rossignoli, Phys. Rev. A 75 (2007) 032350.
\bibitem{Alcaraz} F.C. Alcaraz, A. Saguia, M.S. Sarandy, Phys. Rev. A 70 (2004) 032333.
\end{thebibliography}
\end{document}